\documentclass[aps,preprint]{revtex4}

\usepackage{amssymb}

\usepackage{amsmath}
\usepackage{graphicx}
\usepackage{ulem}
\usepackage{color}

\begin{document}

\title[Spooky van der Waals interactions]{Spooky correlations and unusual
van der Waals forces between gapless and near-gapless molecules}
\author{John F. Dobson}
\affiliation{School of Narural Sciences and Queensland Micro and Nano Technology Centre,
Griffith University, Nathan, Queensland 4111, Australia}
\author{Andreas Savin}
\affiliation{UPMC Sorbonne Universit\'{e}s, Laboratoire de Chimie Th\'{e}orique, 4, place
Jussieu, Case courrier 137,F-75252 Paris Cedex 05, France}
\author{Janos G. Angyan }
\affiliation{Laboratoire de Cristallographie, R\'esonance Magn%
\'{e}tique et Mod\'{e}lisations (CRM2, UMR CNRS 7036)}
\affiliation{Institut Jean Barriol, Universit\'{e} de Lorraine BP %
239, Boulevard des Aiguillettes, 54506 Vandoeuvre-l\`{e}s-Nancy, France}
\author{Ru-Fen Liu}
\affiliation{Institut Jean Barriol, Universit\'{e} de Lorraine, BP %
239, Boulevard des Aiguillettes, 54506 Vandoeuvre-l\`{e}s-Nancy,France}

\begin{abstract}
We consider the zero-temperature van der Waals interaction between two
molecules, each of which has a zero or near-zero electronic gap between a
groundstate and the first excited state, using a toy model molecule (
equilateral H$_{3}$) as an example. We show that the van der Waals energy
between two groundstate molecules falls off as $D^{-3}$ instead of the usual 
$D^{-6}$ dependence, when the molecules are separated by distance $D.\,$\ We
show that this is caused by perfect ''spooky''\ correlation between the two
fluctuating electric dipoles. The phenomenon is related to, but not the same
as, the ''resonant''\ interaction between an electronically excited and a
groundstate molecule introduced by Eisenschitz and London in 1930. It is
also an example of ''type C van der Waals non-additivity''\ recently
introduced by one of us ( Int. J. Quantum Chem. 114, 1157 (2014)). Our toy
molecule H$_{3}$ is not stable, but symmetry considerations suggest that a
similar vdW phenomenon may be observable, despite Jahn-Teller effects, in
molecules with discrete rotational symmetry and broken inversion symmetry,
such as certain metal atom clusters. The motion of the nuclei will need to
be included for a definitive analysis of such cases, however.
\end{abstract}

\maketitle
\affiliation{Micro and Nano Technology Centre, Griffith University, Nathan, Queensland,
Australia}
\affiliation{Sorbonne Universities Pierre et Marie Curie}
\altaffiliation{CNRS}
\affiliation{University of Lorraine, Nancy, France}
\altaffiliation{CNRS}
\affiliation{University of California, Santa Barbara}
\altaffiliation{University of Lorraine, Nancy, France}
\altaffiliation{}


It is well known that a pair of molecules ''$a"$ and $"b"$ in their dimer
groundstate experience an atractive dispersion (van der Waals) interaction
at non-overlapping separations $D$. The dispersion interaction comes from
coupled quantum-fluctuating multipoles. For $D$ much greater than the
molecular sizes, the dipolar $D^{-6}$ interaction term dominates, and is
given by 2nd order perturbation theory as 
\begin{equation}
E{^{(2)}}=-\sum\limits_{J,K}{{\frac{{{{\left| {\left\langle {{0_{a}}{0_{b}}}%
\right| {\frac{{{e^{2}}}}{{{D^{3}}}}}}\left( {{\,\vec{X}_{a}}.{\,\vec{X}%
_{b}-3}}\left( {{\hat{D}.\vec{X}_{a}}}\right) \left( {{\,\hat{D}.\vec{X}_{b}}%
}\right) \right) {\left| {{J_{a}}{K_{b}}}\right\rangle }\right| }^{2}}}}{{%
\left( {{E_{Ja}}-{E_{0a}}}\right) +\left( {{E_{Kb}}-{E_{0b}}}\right) }}}}=-{%
C_{6}}^{ab}{D^{-6}\;\;\;.}  \label{2ndOrderDispersion}
\end{equation}%
Here $\left| 0\right\rangle _{a}$ annd $E_{0a}$ are the many-electron
groundstate and energy of molecule $a.$ and $-\left| e\right| \vec{X}%
_{a}=-\left| e\right| \sum_{i}\vec{r}_{ia}$ is the dipole operator for
molecule $a$, in which the electronic position operator $\vec{r}_{ia}$ of
electron number $i$ in molecule $a$ is measured from the centre of charge of
the constitutent nuclei.\ ${\left| {{J_{a}}}\right\rangle }$,\ $E_{Ja}$ {are
many-electron eigenfunctions and eigenvalues of molecule }${a}$ in isolation{%
, and }${|{{K_{b}>}}}${, }${{E_{Kb}}}$ are those of molecule $b$ in
islolation. $\hat{D}=\vec{D}/D$ is a unit vector pointing between the
molecules.

Eq (\ref{2ndOrderDispersion}) is no longer valid if the separation $D$ is so
large that the time $\tau _{em}=D/c$ for electromagnetic wave propagation
between the molecules exceeds a typical correlation time $\tau _{el}$ of the
intramolecular dipolar fluctuations.

There is also clearly a problem with Eq (\ref{2ndOrderDispersion}) if both
molecules have degenerate electronic groundstates so that there exist
many-electron states $\left| J\right\rangle _{a}$ and $\left| K\right\rangle
_{b}$ such that $E_{Ja}=E_{0a}$ and ${{E_{Kb}}}={{E_{0b}.}}${{\ Then,
provided that the relevant dipole matrix element in (\ref{2ndOrderDispersion}%
) is nonzero, a discrete term in (\ref{2ndOrderDispersion}) has a zero
denominator leading to a dispersion interaction that does not fall off as }}$%
D^{-6}$ (i.e. $C_{6}^{ab}\rightarrow \infty $). Here we will discuss the
consequences of this degenerate situation, starting from the (unrealistic)
toy model of the H$_{3}$ molecule in which the nuclei are constrained to lie
on an equilateral triangle. We will then move on to the possibilities for
observation of similar anomalous vdW interactions between molecules in other
constrained geometries. Finally we will discuss the prospects for
observation of \ such unconventional dispersion forces between real cluster
molecules where Jahn-Teller physics and nuclear motion (pseudo-rotation %
\cite{MDalkaliMetalClustersM3},\cite{HamReDynJahnTeller}) are probably
significant.

In such degenerate situations we show below that there are ''spooky''\
correlations between the fluctuating dipoles on the molecules, correlations
that do not decay with separation $D$, and as a result the dispersion
interaction falls off as $-D^{-3}$ rather than the conventional $-D^{-6}$.

In fact similar unusual van der Waals interactions, falling off with an
unconventional power of separation, have previously been predicted between
extended, low-dimensional nanostructures with a zero HOMO-LUMO gap.
Specifically, for two parallel two-dimensional electron gases separated by
distance $Z$, a sum of $D^{-6}$ atom-atom contributions predicts a vdW
interaction $Z^{-4}$ whereas microscopic theory gives $Z^{-5/2}$ \cite%
{BostromSernelius}. For parallel undoped graphene sheets the conventional
summed result is $E\left( Z\right) \propto -Z^{-4}$ whereas microscopic
theory gives $E\left( Z\right) \propto -Z^{-3}$ or a logarithmically
corrected version of this \cite{Dobson06}\cite{GrGrBeyondRPA}. For the
one-dimensional case of two parallel metallic nanotubes at separation $Z$,
the conventional summed result is $E\left( Z\right) \propto -Z^{-5}$ whereas
more acurate microscopic approaches yield $E\left( Z\right) \propto
-Z^{-2}\left( \ln \left| z\right| \right) ^{-3/2}$ \cite{Chang71},\cite%
{Dobson06},\cite{DrummondNeeds07}\cite{CrossedNanotubesJFD} in the
electromagnetlicaly non-retarded regime. These 2D and 1D nanosystems were
argued \cite{Dobson06},\cite{IJQCTypeABC},\cite{vdWPpointing},\cite%
{HChains2011} to exhibit unconvential vdW powers because of their zero
electronic energy gap and their low dimensionality (limiting the influence
of coulomb screening). In a recent work \cite{IJQCTypeABC}, these
unconventional dispersion power laws were attributed to ''Type-C vdW 
non-additivity''\ arising from the de-localization (hopping) of electrons
between nuclear centres, i.e. to number fluctuations on each centre. 

To explain the last statement it is useful here to summarize the three types of non-pairwise additivity introduced in \cite{IJQCTypeABC}:

{\em Type A} (atomic environment effect). The dispersion interaction
between a single pair of atoms can be expressed in terms of their electric
polarizabilities \cite{LonguetHiggins:65,Zaremba:76}. Type-A non-additivity arises from changes in
the polarizability of an atom because of its confinement inside a molecule
or other structure, arising for example from compression due to Pauli
repulsion from neighboring atoms. \ This means that one cannot simply use
the gas-phase (free-atom) polarizability when computing the vdW energy
contribution of any pair of atoms that are part of larger structures. \ All
of the modern efficient atom-based vdW formalisms (e.g. \cite{Tkatchenko:09},%
\cite{Grimme:10}) allow for this Type-A effect.

{\em Type B} (Coulomb screening by spectator atoms). In the dispersion
interaction between two atoms, an electric field couples the electric
fluctuations on each atom to those of the other, causing correlations
between these fluctuations, correlations whose energy is the dispersion
energy. These intermediary electric fields can be changed (screened) when
they polarize additional atoms, which produce additional induced fields that
affect the original two atoms. \ This results in an interaction between a
collection of more than two atoms, an interaction that is not expressible as
a sum over individual pairs of atoms. This Type-B effect is missing in the
original pairwise atom-based versions of both the popular Grimme\cite%
{Grimme:10} and Tkatchenko-Scheffler \cite{Tkatchenko:09} approaches to the
dispersion energy. \ It is, however, partially included (up to the
three-atom Axilrod-Teller-Muto tern) in the Grimme DFT-D3 approch \cite%
{Ehrlich:11} \cite{Moellmann:14}. The Type-B effects are rather completely
described in the  ''many-body dispersion'' (MBD) method of Tkatchenko and co-workers. \cite%
{Tkatchenko:12}  In this approach each atom is treated as a harmonic
oscillator and the Coulomb interactions between atoms are linearized,
resulting in an exactly soluble model whose correlation energy is the sum of
zero-point energies of collective modes. \ Apart from its harmonic-atom
approximation, the Tkatchenko approach is closely related to the Random
Phase Approximation correlation energy, which can also be written as a sum
of zero-point mode energies \cite{Eshuis:12b}. Significant qualitative
effects of this Type-B physics include a dependence of the asymptotic vdW
energy $E$ on the number $N$ of atoms that is different from the $E\propto
N^{2}$ dependence predicted by summing pairs of atoms \cite{Gobre:13},\cite%
{DiStasioJr:14} \cite{Perdew:12},\cite{Ruzsinszky:12}. (The last two cited
papers used a jellium model for fullerenes and may therefore have included
some Type-C effects: see the next paragraph. Possibly for this reason, the
N-dependence is slightly different from that in the previous two citations).
Type-B physics also causes quantitative differences from pairwise
predictions at all separations, especially for highly polarizable matter %
\cite{Gobre:13}.

{\em Type C}. (Electron hopping / tunneling / number fluctuation effect). \
In this scenario, large fluctuating electrical moments can occur via the
movement of electrons between atomic centers, with a consequent contribution
to the dispersion interaction. This effect is not possible, for example, in
theories \cite{Grimme:10}, \cite{Tkatchenko:09}, \cite{Tkatchenko:12} that
assign an electron uniquely to one atom or another. \ The consequences of
Type-C physics for the dispersion energy $E$ include exponents $p$, in the
asymptotic decay $E\approx -AD^{-p}$ with separation $D$, that differ from
the value predicted by summing over atom pairs \cite{Dobson06}. \ This
Type-C physics rarely occurs alone: it can be heavily modified by Type-B
electric screening of the large electric moments that Type-C physics, on its
own, would create. \ In particular, in three-dimensional metals the (Type-B)
Coulomb screening is complete, and the Type-C physics makes no {\em %
qualitative} difference. \ On the other hand, Coulomb screening is
incomplete in lower-dimensional zero-gap systems, and there the Type-C
effect causes unusual asymptotic vdW power laws. For example, the \emph{asymptotic}
interaction $E$ between two parallel undoped graphene sheets at low
temperature, separated by distance $D$, is predicted by high-level theory to
be of form $E\approx -A_{3}D^{-3}$\cite{Dobson06}, whereas the prediction of
atom-pair summation is of form $E\approx -A_{4}D^{-4}$. \ At small
separations $D<5\,nm$ the gapless transitions that gave rise to the above
unusual asymptotic interaction constitute only a small aspect of the vdW
interaction, and Type-C effects are not important in graphene at these
separations. \ For two 2-dimensional metals the result is $E\approx
-A_{5/2}D^{-5/2}.%
$\cite{BostromSernelius}. For two
parallel conducting one-dimesional wires, , the result of high-level theory
is $E\approx -A_{2}D^{-2}(\ln D/D_{0})^{-3/2}$ compared with
 $E\approx -A_{5}D^{-5}$ from pairwise atomic summation 
\cite{Chang71},\cite{Dobson06},\cite{DrummondNeeds07}. These results suggest that the
Type-C effects are more significant in lower dimensionality. From this way
of thinking, one expects that there can be very strong type-C effects
between \ two gapless ''zero-dimensional'' systems,  i.e. two
small molecules, each with a degenerate groundstate. \ The toy model of the H%
$_{3}$ dimer considered in the present work is a case in point.

An attractive feature of the toy model of the equilateral H$_{3}$ molecule
discussed below is that one can easily isolate the Type-C (electron
inter-atom hopping) effects from Type-B effects (polarization of electrons
on individual atoms). This is simply achieved by a Configuration Interaction
energy calculation for a pair of interacting H$_{3}$ molecules, in which the
one-electron basis is restricted to one s-state on each hydrogen atom. \
Then dynamic distortion / polarization of the electronic charge 
distribution within one atom is not possible, and flucuating dipole moments
are generated solely via hopping of electrons between s states on different
atoms. Some Type-B effects can later be introduced by including p states in
the basis, since a superposition of an s and a p state exhibits polarization
of the electronic cloud on a given H atom. This is done in the Sections below.

\section{A toy model of spooky dipolar interactions: equilateral H$_{3}$}

To observe the spooky dipolar correlations and $-D^{-3}$ interaction in the
electronic groundstate of a moleular pair as proposed above, the discussion
following Eq (\ref{2ndOrderDispersion}) suggests that one needs to find a
molecule with two strictly degenerate groundstates that are coupled by the
electric dipole operator \ $\ $Such molecules are not easy to find. The
idealized case of a strictly equilateral H$_{3}$ molecule is one such case,
as we show below. However, a literature search suggests that the H$_{3}$
molecule is not stable in its electronic groundstate, and previous work
found that the theoretical H$_{3}$ conformation of minimum electronic
groundstate energy would be a linear geometry, not a triangle. \ In general,
even where a candidate degenerate molecule for these effects is stable, one
can exepect a Jahn-Teller effect to occur, whereby a fully isolated molecule
will distort geometrically. We discuss these effects later on in this paper.
For now we artificially hold the protons in the equilateral triangle
conformation. The toy model that then results is very informative for
present purposes. \ We can adjust the distance between the protons withn a
molecule in order to explore regimes of weak and strong orbital overlap.

\subsection{Minimal-basis, independent-electron model of a single
equilateral H$_{3}$}

We first study the simplest possible version capable of electronic dipolar
excitations, namely three independent electrons moving in the Coulomb
potential due to the nuclei, with a basis consisting of only a single $s$
state on each nucleus. Thus a distortion of the charge cloud on each nucleus
is not possible, and the dipolar fluctuations that lead to the dispersion
interaction between two such idealized molecules arise from hopping of the
electrons between the localized $s$ states on the different protons of one
molecule. (This will lead to the the pure ''type-C non-additivity
''phenomenon in dispersion interactions between such molecules, as proposed
in \cite{IJQCTypeABC}). \ We first show that this non-interacting
three-electron model for a single H$_{3}$ molecule has 2 exactly degenerate
electronic groundstates for each allowed spin configuration. For each spin
configuration we further show that the two groundstates are coupled by the
dipole moment operator. We will then use a limited Configuration Interaction
(CI)\ approach to a pair of H$_{3}$ molecules, leading to the spooky $%
-D^{-3} $ inter-interaction as described in general above. In the following
section we will then show, by symmetry arguments and limited CI\
calculations, that these conclusions survive even when electron-electron
interactions are re-introduced\ and a larger basis is used. \ In a further
Section we confirm these conclusions via a larger CI\ calculation using the
package MOLPRO.

\begin{figure}
\includegraphics[width=1.0\linewidth]{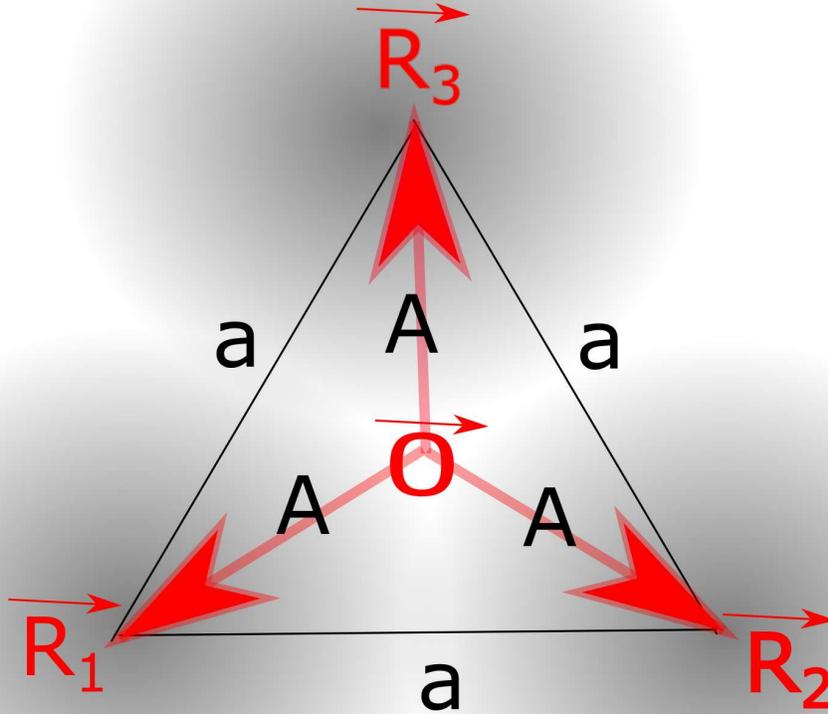}%
\caption{Geometry of equilateral H$_3$}
\label{Fig1}
\end{figure}

We label the three protons $i=1,2,3$ (see Fg 1) . The 6 localized
1-electron $s$ states are denoted $\left| i\uparrow \right\rangle $, $\left|
i\downarrow \right\rangle $ in Dirac notation. The 1-electron Hamiltonian
for a spin-up electron is of form%
\begin{equation}
H_{\uparrow }^{1\;electron}=\left( 
\begin{array}{ccc}
\varepsilon & -t & -t \\ 
-t & \varepsilon & -t \\ 
-t & -t & \varepsilon%
\end{array}%
\right)  \label{1ElH}
\end{equation}%
where $\varepsilon $ is the 1s energy and $t$ is a ''hopping''\ matrix
element of the hamiltonian between any two sites. $t$ is usually positive
except possibly when the internuclear distance is chosen to be extremely
small. The matrix element of the position operator $\vec{r}$ is 
\begin{equation}
\left\langle i\uparrow \right| \vec{r}\left| j\uparrow \right\rangle
=\left\{ 
\begin{array}{c}
\left( \vec{R}_{i}+\vec{R}_{j}\right) \alpha /2.\ \ i\neq j \\ 
\vec{R}_{i},\ \ \ \ \ \ \ i=j%
\end{array}%
\right.  \label{rMatrixEltLocBasis}
\end{equation}%
where $\vec{R}_{i}$ is the location of the $i^{th}$ proton (see Fig 1 for
molecular geometry and labelling). The neighbor overlap element $\alpha $ in
(\ref{rMatrixEltLocBasis}) is real and is the same for all neighbor pairs,
by symmetry. 

We consider the spin-up case for definiteness. A convenient, normalized,
maximally symmetric set of 1-electron eigenfunctions of (\ref{1ElH}) are the
three Bloch states 
\begin{eqnarray}
\left| g\uparrow \right\rangle &=&\frac{1}{\sqrt{3\left( 1+2\alpha \right) }}%
\left( \left| 1\uparrow \right\rangle +\left| 2\uparrow \right\rangle
+\left| 3\uparrow \right\rangle \right) ,\ \ \ \varepsilon _{g\uparrow
}=\varepsilon -2t  \notag \\
\left| +\uparrow \right\rangle &=&\frac{1}{\sqrt{3\left( 1-\alpha \right) }}%
\left( e^{-i2\pi /3}\left| 1\uparrow \right\rangle +e^{-i4\pi /3}\left|
2\uparrow \right\rangle +\left| 3\uparrow \right\rangle \right)
,\;\;\;\varepsilon _{+\uparrow }=\varepsilon +t  \notag \\
\left| -\uparrow \right\rangle &=&\frac{1}{\sqrt{3\left( 1-\alpha \right) }}%
,\ \ \left( e^{+i2\pi /3}\left| 1\uparrow \right\rangle +e^{+i4\pi /3}\left|
2\uparrow \right\rangle +\left| 3\uparrow \right\rangle \right)
,\;\;\varepsilon _{-\uparrow }=\varepsilon +t  \label{BlochStates}
\end{eqnarray}%
where normalization has been ensured by introducing the overlap matrix
elements%
\begin{equation*}
\left\langle 1\uparrow \right. \left| 1\uparrow \right\rangle =1,\ \ \
\alpha =\left\langle 1\uparrow \right. \left| 2\uparrow \right\rangle
\end{equation*}%
The states $\left| +\uparrow \right\rangle $, $\left| -\uparrow
\right\rangle $ describe an electron circulating (hopping) clockwise or
anti-clockwise round the triangular molecule, respectively. \ The three
Bloch states are eigenfunctions of the 120$^{0}$ rotation operator $\mathcal{%
\hat{R}}_{120}$, with eigenvalues $1,e^{i2\pi /3}$ and $e^{-i2\pi /3}$
respectively. The $+$ and $-$ $\ $states are related by the time reversal
operation $\hat{T}$ (complex conjugation with spin not included):\ $\ \left|
-\uparrow \right\rangle =\hat{T}\left| +\uparrow \right\rangle $ \ Note that
the Hamiltonian commutes with $\mathcal{\hat{R}}_{120}$ and $\hat{T}$, a
property which will survive in the more sophisticated interacting models of H%
$_{3}$ to be discussed below. The three Bloch states span the one-body space
in the present limited basis.

The matrix elements of the electron position operator $\vec{r}$ between the
Bloch states are, from (\ref{rMatrixEltLocBasis}) and (\ref{BlochStates})%
\begin{eqnarray}
\left\langle +\right| \vec{r}\left| -\right\rangle &=&-\frac{1}{2}A(i\hat{x}-%
\hat{y})=\left\langle -\right| \vec{r}\left| +\right\rangle ^{\ast }\ \ \ 
\label{BlochMatrixEltsr} \\
\ \left\langle +\right| \vec{r}\left| +\right\rangle \ \ &=&\ \left\langle
-\right| \vec{r}\left| -\right\rangle \ \ =\left\langle g\right| \vec{r}%
\left| g\right\rangle =\vec{0}  \label{BlochMatrixEltsrB}
\end{eqnarray}%
provided that $\vec{r}$ is measured from the centroid of the proton
triangle. Here $A$ is the distance of each proton from the centroid of the
triangle, so that the proton-proton distance is $a = \sqrt{3}A.$ (See Fig 1)

Two independent, exactly degenerate 3-electron determinantal groundstates $%
\left| G^{+}\right\rangle $ and $\left| G^{-}\right\rangle $ each with total
spin projection +$\hbar /2$ are made by doubly occupying ($\uparrow
,\downarrow $) the zero Bloch state $g$ while occupying either the + or -
Bloch state with an $\uparrow $ electron: 
\begin{equation}
\left| G^{+}\right\rangle =\hat{c}_{g\uparrow }^{\dagger }\hat{c}%
_{g\downarrow }^{\dagger }\hat{c}_{+\uparrow }^{\dagger }\left|
0\right\rangle ,\ \ \left| G^{-}\right\rangle =\hat{c}_{g\uparrow }^{\dagger
}\hat{c}_{g\downarrow }^{\dagger }\hat{c}_{-\uparrow }^{\dagger }\left|
0\right\rangle  \label{GAndFStatesForIndepEH3}
\end{equation}%
where the repeated creation operators $\hat{c}$, acting on the vacuum $%
\left| 0\right\rangle $, generate determinantal states formed from
one-electron Bloch orbitals such as $\phi _{+\uparrow }\left( \vec{r}\right)
=\left\langle \vec{r}\right. \left| +\uparrow \right\rangle $, with the
correct Fermionic antisymmetry. \ See Fig 2

\begin{figure}
\includegraphics[width=1.0\linewidth]{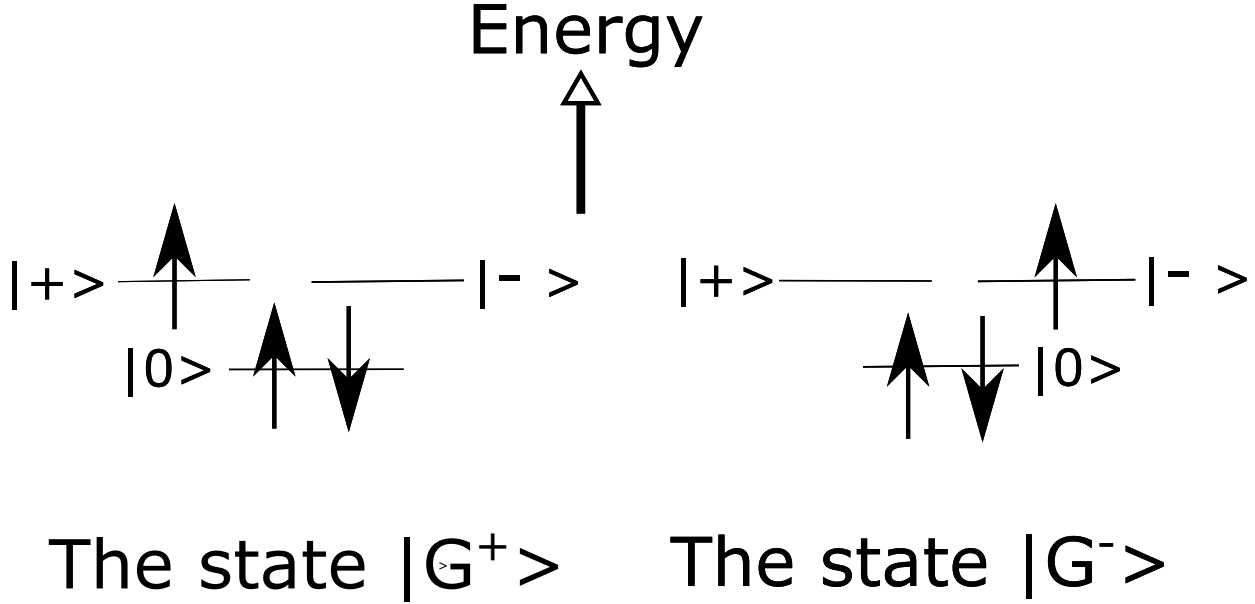}%
\caption{Two degenerate independent-electron grounstates based on clockwise and anticlockwise Bloch orbitals}
\label{Fig2}
\end{figure}
\section{Equilateral H$_{3}-$H$_{3}$ interaction}

We consider two H$_{3}$ molecules labelled ''a'' and ''b''. For simplicity
we restrict attention to the ''facing directly opposite'' geometry where the
centroids of the molecules are separated by the displacement $\vec{D}=D\hat{z%
},$ and the plane of each molecule is parallel to the xy plane, with the
protons aligned. Then the dipolar inter-molecular coupling Hamiltonian (see %
\ref{2ndOrderDispersion}) simplifies to 
\begin{equation}
\delta H_{ab}=e^{2}D^{-3}\vec{X}_{a}.\vec{X}_{b},\;\;\;\;\;\;\;\;\vec{X}%
_{a}=\sum_{i=1}^{3}\vec{r}_{ia}  \label{DipoleHFacing}
\end{equation}%
where $\vec{X}_{a}$ is the total position operator for the electrons in
molecule $a$. We evaluate the energy of the 6-electron, two- molecule system
in a limited Configuration Interaction approach, keping only the two
degenerate 3-electron groundstates on each molecule, leading to a fourfold
product-basis set 
\begin{equation}
\left| G^{+}\right\rangle _{a}\left| G^{+}\right\rangle _{b},\;\;\;\left|
G^{+}\right\rangle _{a}\left| G^{-}\right\rangle _{b},\;\;\;\;\left|
G^{-}\right\rangle _{a}\left| G^{+}\right\rangle _{b},\;\;\;\left|
G^{-}\right\rangle _{a}\left| G^{-}\right\rangle _{b}  \label{MinimalCIBasis}
\end{equation}%
or more compactly$.\left| ++\right\rangle ,\left| +-\right\rangle ,\left|
-+\right\rangle ,\left| --\right\rangle .$

The intermolecular coupling matrix elements are

\begin{eqnarray*}
\left\langle ++\right| \delta H_{ab}\left| --\right\rangle &=&\frac{e^{2}}{%
D^{3}}\left\langle G_{a}^{+}G_{b}^{+}\right| \vec{X}_{a}.\vec{X}_{b}\left|
G_{a}^{-}G_{b}^{-}\right\rangle \\
&=&\frac{e^{2}}{D^{3}} ( \left\langle +\right| x\left| -\right\rangle
_{a}\left\langle +\right| x^{\prime }\left| -\right\rangle _{b}+\left\langle
+\right| y\left| -\right\rangle _{a}\left\langle +\right| y^{\prime }\left|
-\right\rangle _{b} ) \\
\ &=&\frac{e^{2}}{D^{3}}\left[ \frac{-1}{2}A(i\hat{x}-\hat{y})_{x}\frac{-1}{2%
}A(i\hat{x}-\hat{y})_{x}\right] +\frac{e^{2}}{D^{3}}\left[ \frac{-1}{2}A(i%
\hat{x}-\hat{y})_{y}\frac{-1}{2}A(i\hat{x}-\hat{y})_{y}\right] \\
&=&\frac{e^{2}}{4D^{3}}\left( -A^{2}\right) +\frac{e^{2}}{4D^{3}}\left(
A^{2}\right) =0\;
\end{eqnarray*}

The vanishing of this matrix element can be understood because the
two-molecule states $\left| G_{a}^{+}G_{b}^{+}\right\rangle $ and $\left|
G_{a}^{-}G_{b}^{-}\right\rangle $ have different eigenvalues of the total
(discrete) angular momentum variable $\mathcal{L}_{120}$ that generates the
(discrete) rotation operators $1,\mathcal{R}_{120},\mathcal{R}_{120}^{2}$,
whereas the scalar operator $\vec{X}_{1}.\vec{X}_{2}$ is fully rotationally
invariant and so conserves $\mathcal{L}_{120}$.

The diagonal elements are also zero, $\left\langle ++\right\vert \delta
H_{ab}\left\vert ++\right\rangle =\left\langle --\right\vert \delta
H_{ab}\left\vert --\right\rangle =\left\langle +-\right\vert \delta
H_{ab}\left\vert +-\right\rangle =$ $\left\langle -+\right\vert \delta
H_{ab}\left\vert -+\right\rangle $ but the following cross-term is nonzero

\begin{eqnarray}
\left\langle +-\right| \delta H_{ab}\left| -+\right\rangle &=&\frac{e^{2}}{%
D^{3}}\left\langle G_{1}^{+}G_{2}^{-}\right| \vec{X}_{1}.\vec{X}_{2}\left|
G^{-}G_{2}^{+}\right\rangle =\frac{e^{2}}{D^{3}}\left( \left\langle +\right|
x\left| -\right\rangle _{1}\left\langle -\right| x^{\prime }\left|
+\right\rangle _{2}+\left\langle +\right| y\left| -\right\rangle
_{1}\left\langle -\right| y^{\prime }\left| +\right\rangle _{2}\right) 
\notag \\
&=&\frac{e^{2}}{D^{3}}\left[ \frac{-1}{2}A(i\hat{x}-\hat{y})_{x}\frac{-1}{2}%
A(i\hat{x}-\hat{y})_{x}^{\ast }\right] +\frac{e^{2}}{D^{3}}\left[ \frac{-1}{2%
}A(i\hat{x}-\hat{y})_{y}\frac{-1}{2}A(i\hat{x}-\hat{y})_{y}^{\ast }\right] 
\notag \\
&=&\frac{e^{2}}{4D^{3}}A^{2}+\frac{e^{2}}{4D^{3}}A^{2}=\frac{e^{2}A^{2}}{%
2D^{3}}  \label{NonZeroABBlochElement}
\end{eqnarray}

We measure energies relative the groundstate energy of two independent H$%
_{3} $ molecules, and the CI hamiltonian then becomes extremely simple (with
the states ordered as in Eq (\ref{MinimalCIBasis})):%
\begin{equation}
\hat{H}_{ab}=\left( 
\begin{array}{cccc}
0 & 0 & 0 & 0 \\ 
0 & 0 & \mu D^{-3} & 0 \\ 
0 & \mu ^{\ast }D^{-3} & 0 & 0 \\ 
0 & 0 & 0 & 0%
\end{array}%
\right) ,\;\;\mu =e^{2}A^{2}/2  \label{Hab}
\end{equation}%
Two independent bi-molecular states diagonalizing this Hamiltonian are%
\begin{eqnarray}
\left| \Phi _{e}\right\rangle &=&\frac{1}{\sqrt{2}}\left( \left|
+-\right\rangle -\left| +-\right\rangle \right) ,\ \ E_{e}=-\frac{e^{2}A^{2}%
}{2D^{3}}  \label{Phie} \\
\left| \Phi _{f}\right\rangle &=&\frac{1}{\sqrt{2}}\left( \left|
+-\right\rangle +\left| +-\right\rangle \right) ,\ \ E_{f}=+\frac{e^{2}A^{2}%
}{2D^{3}}  \notag
\end{eqnarray}%
The state $\left| \Phi _{e}\right\rangle $ is the groundstate of the H$_{3}-$%
H$_{3}$ system, and its energy $-e^{2}A^{2}/\left( 2D^{3}\right) $ falls off
with separation $D\,\ $as $D^{-3},$ instead of the usual dispersion (vdW)
energy, which varies as $D^{-6}$. This, with generalization to
interacting electrons and a larger single-electron basis as discussed below,
is a principal result of the present work.

\section{Understanding the spooky state of the equilateral H$_{3}$ dimer as
a pair of fluctuating but perfectly correlated electric dipoles}

The correlated state (\ref{Phie}) is more easily understood by re-expressing
it in terms of real 1-electron orbitals $\left| up\right\rangle $, $\left|
down\right\rangle $ with overt dipole moments, orbitals that are less
symmetric than the clockwise and anticlockwise Bloch orbitals $\left|
+\right\rangle $. $\left| -\right\rangle $ introduced in Eq (\ref%
{BlochStates}). We make the (non-unique) choice

\begin{equation*}
\left| up\right\rangle =\frac{1}{\sqrt{2}}\left( \left| +\right\rangle
+\;\left| -\right\rangle \right) ,\;\;\left| down\right\rangle =\frac{1}{%
\sqrt{2}i}\left( \left| +\right\rangle -\;\left| -\right\rangle \right)
\end{equation*}%
The 1-electron charge densities from these two states have dipole moments%
\begin{equation}
\pm \frac{1}{2}\left| e\right| A/\left( 1-\alpha \right) \;\;\;,
\label{DipoleMoments}
\end{equation}%
pointing towards and away from nucleus \# 3 of the H$_{3}$ triangle
respectively. In the 3-electron groundstate manifold of H$_{3}$, the
doubly-occupied lowest-lying orbital $\left| g\right\rangle $ contributes no
electric dipole moment and so we can generate 2 alternative 3-electron
groundstates $\left| G^{up}\right\rangle ,~\left| G^{down}\right\rangle $
with the same dipole moments as in Eq (\ref{DipoleMoments}): $\ $%
\begin{eqnarray}
\left| G^{up}\right\rangle &=&\frac{1}{\sqrt{2}}\left( \left|
G^{+}\right\rangle +\left| G^{-}\right\rangle \right) =\hat{c}_{g\uparrow
}^{\dagger }\hat{c}_{g\downarrow }^{\dagger }\hat{c}_{up\uparrow }^{\dagger
}\left| 0\right\rangle \   \label{Gup} \\
\left| G^{down}\right\rangle &=&\frac{1}{\sqrt{2}i}\left( \left|
G^{+}\right\rangle -\left| G^{-}\right\rangle \right) =\hat{c}_{g\uparrow
}^{\dagger }\hat{c}_{g\downarrow }^{\dagger }\hat{c}_{down\uparrow
}^{\dagger }\left| 0\right\rangle  \label{Gdown} \\
&\therefore &\left| G^{\pm }\right\rangle =\frac{1}{\sqrt{2}}\left( \left|
G^{up}\right\rangle \pm i\left| G^{down}\right\rangle \right)
\label{GpmfromGupdown}
\end{eqnarray}%
We can use (\ref{GpmfromGupdown}) to write the correlated groundstate from (%
\ref{Phie}) in the form%
\begin{equation}
\left| \Phi _{e}\right\rangle =i\left( \left| G^{up}\right\rangle _{a}\left|
G^{down}\right\rangle _{b}-\left| G^{down}\right\rangle _{a}\left|
G^{up}\right\rangle _{b}\right)  \label{AnticorrelatedDipolesState}
\end{equation}%
This exbibits $\left| \Phi _{e}\right\rangle $ as a state with perfect
anticorrelation between the electric dipole moments on the two H$_{3}$
molecules.: when one is ''up''\ the other is ''down'', and vice versa \ This
perfect correlation does not decay with intermolecular distance $D$, since
the coefficents $\pm i$ in the superposition (\ref%
{AnticorrelatedDipolesState}) are $D$-independent. This means that, although
the $D^{-3}$ decay of the vdW interaction here is same as for two fixed
dipoles, this is different from that case because the interaction can be
repulsive for fixed dipoles depending on orientation, whereas the present
effect is always attractive in the dimer groundstate. It is a true van der
Waals interaction. It can also be compared with the non-decaying ''spooky''\
correlations between spins or between photons in the study of quantum
computing situations. The difference here is that the correlated entities
are electric dipoles rather than electron spins or photons.

It is also instructive, for later use, to write the two-molecule CI\
Hamiltian in the following basis of \ electric dipole states 
\begin{equation*}
\left| G^{up}\right\rangle _{a}\left| G^{up}\right\rangle _{b},\,\;\;\left|
G^{up}\right\rangle _{a}\left| G^{down}\right\rangle _{b},\;\;\left|
G^{down}\right\rangle _{a}\left| G^{up}\right\rangle _{b},\;\;\left|
G^{down}\right\rangle _{a}\left| G^{down}\right\rangle _{b}
\end{equation*}%
giving 
\begin{equation}
H_{ab}^{\prime }=\left( 
\begin{array}{cccc}
\frac{1}{2}\frac{\mu }{D^{3}} & 0 & 0 & \frac{1}{2}\frac{\mu }{D^{3}} \\ 
0 & -\frac{1}{2}\frac{\mu }{D^{3}} & \frac{1}{2}\frac{\mu }{D^{3}} & 0 \\ 
0 & \frac{1}{2}\frac{\mu }{D^{3}} & -\frac{1}{2}\frac{\mu }{D^{3}} & 0 \\ 
\frac{1}{2}\frac{\mu }{D^{3}} & 0 & 0 & \frac{1}{2}\frac{\mu }{D^{3}}%
\end{array}%
\right)  \label{HPrimeab}
\end{equation}%
where $\mu =e^{2}A^{2}/2$. The hamiltonian matrix (\ref{HPrimeab}) naturally
has the same eigenvalues $0,0,-\frac{\mu }{D^{3}},\frac{\mu }{D^{3}}$ \ as
the original matrix (\ref{Hab}) that used the +,- basis. 

Our analysis above is quite consistent with the treatment of "quantum electrical dipoles" given by Allen, Abanov and Requist \cite{Allen05}. $\allowbreak $

\section{Minimal-basis analysis of 3-electron states in H$_{3}$ including
on-site repulsion}

When the electron-electron interaction is included in the model of a single
equilateral H$_{3}$, the interacting hamiltonian still has invariance under
the 120-degree rotation operator $\mathcal{R}_{120}$ and also under time
reversal $\hat{T}$. Therefore, just as for the independent-electron model
above, we expect that there will be a degenerate pair of 3-electron states $%
\left| G^{+}\right\rangle $ and $\left| G^{-}\right\rangle $ that are
analogous to non-interacting states defined in Eq (\ref%
{GAndFStatesForIndepEH3}). In particular they differ by a time reversal and
are eigenfunctions $\mathcal{R}_{120}$ with eigenvalues $\exp \left( \pm
i2\pi /3\right) $. (Note that, since $\mathcal{R}_{120}^{3}=1$, the only
possible eigenvalues of $\mathcal{R}_{120}$ are the three complex cube roots
of 1). The matrix element $\left\langle G_{a}^{+}G_{b}^{+}\right| \vec{X}%
_{a}.\vec{X}_{b}\left| G_{a}^{-}G_{b}^{-}\right\rangle $ will therefore
still be zero by the symmetry argument given above, and we expect $%
\left\langle G_{a}^{+}G_{b}^{-}\right| \vec{X}_{a}.\vec{X}_{b}\left|
G_{a}^{-}G_{b}^{+}\right\rangle $ will still be non-zero. \ The question is
whether these two degenerate states are still groundstates. \ We have set up
a $9\times 9$ spin-restricted Hamiltonian matrix including an on-site
repulsion energy $U$ as well as hopping elements $-t$. \ At fixed on-site
repulsion $U$, we find that two time-reversed degenerate states $\left|
G^{+}\right\rangle $ and $\left| G^{-}\right\rangle $ remain the
groundstates for all positive $t$ except for $t=0$ exactly. For $t=0$ the
three exactly degenerate groundstates have no double occupation of any
proton site. We also verified explicitly that 
\begin{equation*}
\left\langle G_{a}^{+}G_{b}^{+}\right| \vec{X}_{a}.\vec{X}_{b}\left|
G_{a}^{-}G_{b}^{-}\right\rangle =0,\ \ \ \ \left\langle
G_{a}^{+}G_{b}^{-}\right| \vec{X}_{a}.\vec{X}_{b}\left|
G_{a}^{-}G_{b}^{+}\right\rangle \neq 0
\end{equation*}%
However, as $t$ decreases towards zero, the gap from the groundstate doublet
of each H$_{3}$ unit, to the next state, goes towards zero. Thus for small $%
t $, corresponding to a uniformly stretched $H_{3}$ triangle, our two-state
limited CI\ analysis of the H$_{3}-$H$_{3}$ becomes invalid. This means that
when seeking real systems that exhibit the unusual vdW interaction (\ref%
{Phie}), it would be best to look at (e.g.) clusters of metal atoms that
favor hopping of electrons between nuclei, as suggested by the metallic
nature of bulk metals.

\subsection{Larger--basis analysis of interacting 3-electron states in H$%
_{3} $}

We also performed Full CI, limted-basis calculations using MOLPRO, to study the equilateral H$_{3}$ dimer system with a
larger basis and all the electron-electron interactions, reaching the same
qualitative conclusions as above. \ This calculation allows static and
dynamic distortions of the electron density on each proton (type-B
nonadditive vdW effect \cite{IJQCTypeABC}), as well as the
previously-considered ''Type-C''\ effects due to hopping of electrons
between the protons. \ Figure 3 shows the quantity $D^{3}\Delta E$ versus $%
H_{3}-H_{3}$ separation $D$, where $\Delta E=E\left( D\right) -E\left(
\infty \right) $ is the binding energy of the $H_{3}-H_{3}$ dimer. Results
are shown for the T1.2 symmetry, which gives the dimer groundstate at all
the $D$ values considered. \ Two FCI calculations were performed, one with a
1s-only basis and one with p orbitals in the basis as well. \ In each case
the curve becomes flat at larger separations, indicating that $\Delta
E\propto -D^{-3}$ as predicted on symmetry grounds by the above theory.

\begin{figure}
\includegraphics[width=1.0\linewidth]{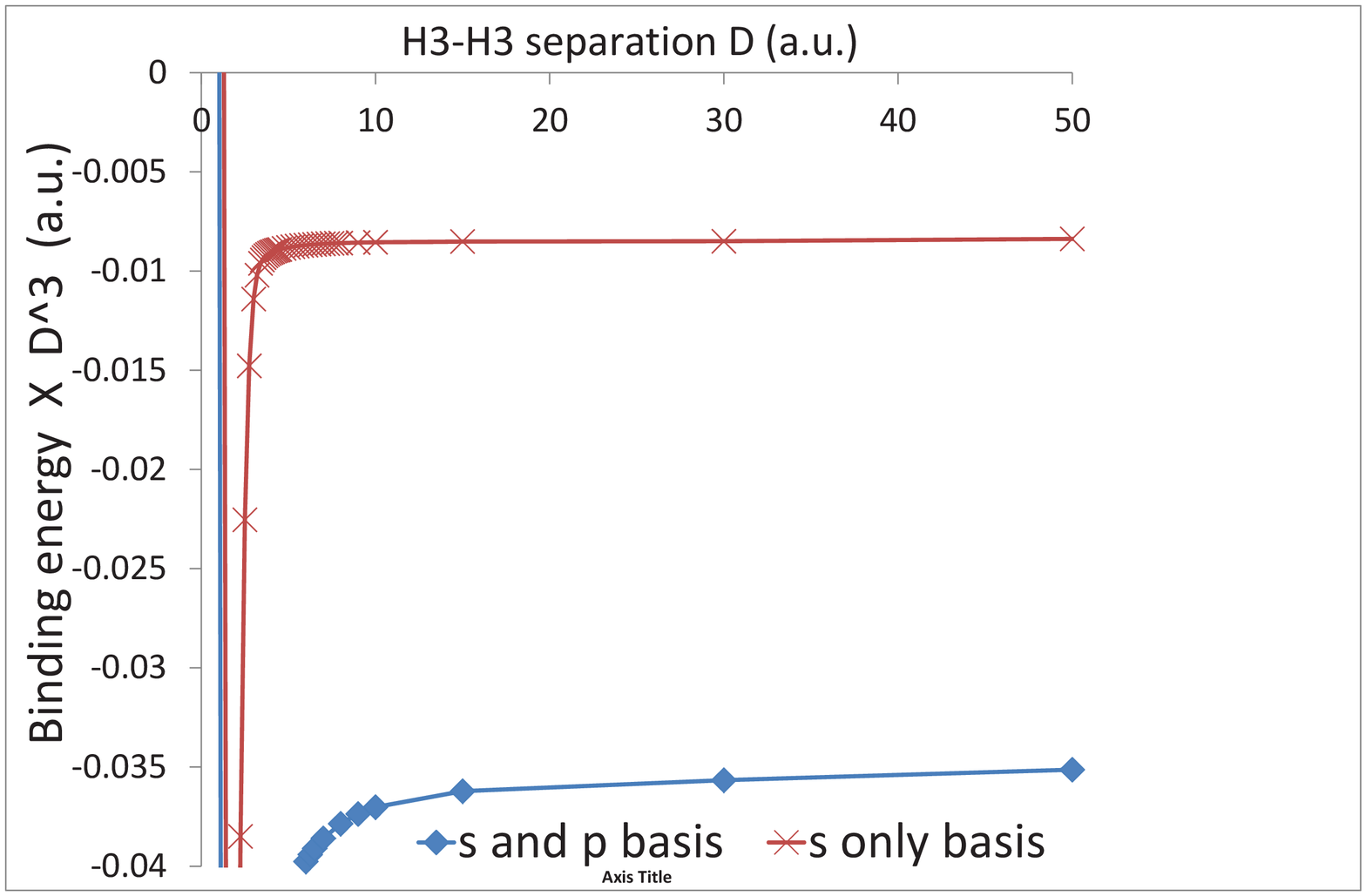}%
\caption{Resuts of FCI calculations demonstrating that the interaction energy $\Delta E$ of two H$_3$ molecules falls off as 
$\Delta E \approx -C_3 D^{-3}$, 
both with an s-only basis and with an s+p basis.  We plot the quantity $X=D^{3} \Delta E$ vs. H$_3$ - H$_3$ separation $D$.   $X$ will be independent of $D$ if the $\Delta E \approx -C_3 D^{-3}$ power law holds. Electron-electron interactions are retained within and between molecules in both cases.  Crosses: s basis only.  Diamonds: s+p basis}
\label{Fig1}
\end{figure}

Interestingly, the \ $D^{-3}$ interaction is stronger with the $p$ orbitals
included, suggesting that the dynamic distortions of the charge cloud on
each atom (Type-B effect \cite{IJQCTypeABC}) are assisting rather than
hindering the type-C (inter-atom hopping) polarizability, in the present geometry.

Elsewhere we plan to present further Full CI data for larger bases, and also for the near-degenerate Jahn-Teller-distorted (isosceles) cases introduced below.

\section{Static Jahn-Teller distortions}

If the nuclear positions of a single isolated molecule are allowed to relax,
the exact electronic degeneracy proposed here can lead to an energy-lowering
Jahn-Teller distortion \cite{JahnTeller}, causing broken rotational symmetry
of the proton configuration in the groundstate. The electronic eigenstates
proposed above will then be replaced by non-degenerate states with a dipole
moment. This completely changes the situation for H$_{3},$ which in fact is
believed to be unstable\ in its electronic groundstate, and if it were
stable, would Jahn-Teller distort continuously to a linear configuration.
Thus our example of equilateral H$_{3}$ is just a toy model requiring an
external agent to hold the nuclear positions fixed. However in cases where
the nuclear framework of a candidate cluster molecule is sufficiently rigid,
such distortions will be small and will introduce only a small energy gap $%
\varepsilon _{g}$ \ Candidate systems for this situation might include
transition or rare earth metal clusters where the observed enhanced rigidity
of the bulk metal (comapred to the simple s-p metals) suggests that
directional bonds may provide the needed structural rigidity, while the
hopping $t$ tends to dominate on-site repulsion $U$ $\ $(leading to full
metallic behavior in the limit of a large number of atoms). 
Metal atom clusters are only one possibility, however, and one could
envisage many other possibilities for stiff structures with the required
symmetry properties, based for example on the rigidity of graphene hexagons.

To commence exloration of the effects of the Jahn-Teller phenomenon, we
therefore now investigate the simplest model with a static distortion,
namely isosceles H$_{3}$ with frozen nuclear positions. \ We will show that
spooky vdW correlations can still occur in a significant subasymptotic
regime of separations provided that the distortion-induced energy gap $%
\varepsilon _{g}$ is small enough.

\section{Triangular H$_{3}$ with a weak isosceles distortion}

We consider an isosceles triangle of protons with base $b$ and height $h$.
with protons at positions $\vec{R}_{1}=-b\hat{x}/2-h\hat{y}/3$, $\ R_{2}=+b%
\hat{x}/2-h\hat{y}/3$, $\ \ \vec{R}_{3}=\frac{2}{3}h\hat{y}$. (measured
relative to the centre of charge O of the protons, at height $h/3$ above the
base. See Fig.4) \ Initially we work again in the independent-electron
model, with 3 electrons moving between the stationary protons.

\begin{figure}
\includegraphics[width=1.0\linewidth]{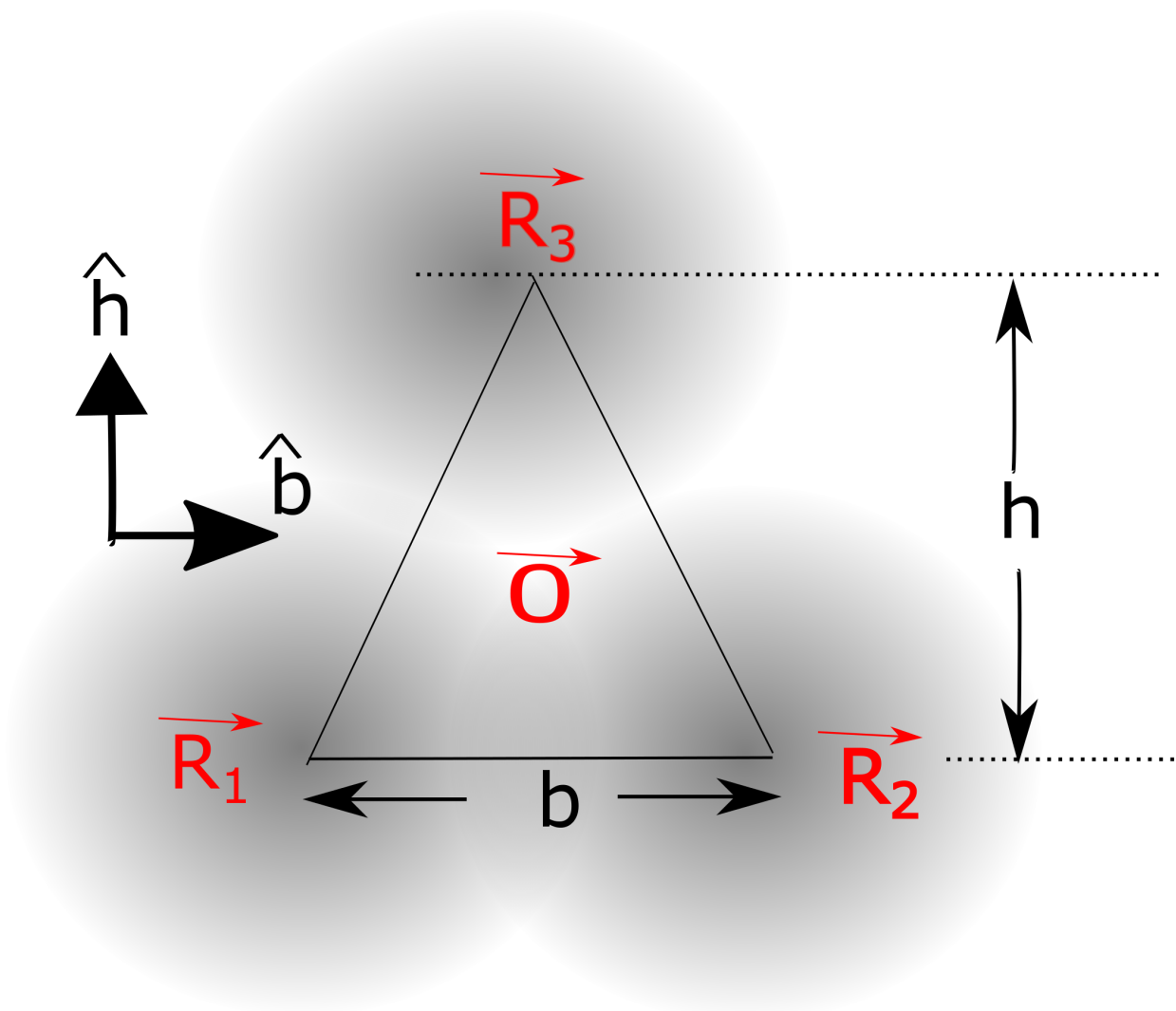}%
\caption{Geometry of isosceles H$_3$}
\label{Fig4}
\end{figure}

The reduction of the hopping matrix elements involving the more distant atom
\#3 leads to a one-electron hamiltonian in the minimal localised atomic
basis $\left| 1\right\rangle $, $\left| 2\right\rangle $, $\left|
3\right\rangle $, centered on the protons, as follows 
\begin{equation}
\hat{H}=\varepsilon +\left( 
\begin{array}{ccc}
0 & -t & -t^{\prime } \\ 
-t & 0 & -t^{\prime } \\ 
-t^{\prime } & -t^{\prime } & -\Delta%
\end{array}%
\right)  \label{IsoscelesH1E1}
\end{equation}

where $t^{\prime }$ is the hopping element to/from the apex, atom \#3 (see
fig. 4) and $\Delta $ is the additional energy of an electron on proton 3 as
a result of orbtal compression in the groundstate. Here for an acute
isosceles triangle we expect $\Delta <0$ and $0<$ $t^{\prime }<t$ so that $%
\delta \equiv t^{\prime }-t<0$ \ For an obtuse isosceles triangle, $\Delta
>0 $ and $\delta >0$. The eigenvalues of (\ref{IsoscelesH1E1}) are%
\begin{eqnarray}
\;\varepsilon ^{\left( 1\right) } &=&\varepsilon +\frac{1}{2}\Delta -\frac{1%
}{2}t-\frac{1}{2}\sqrt{\left( t+\Delta \right) ^{2}+8(t^{\prime })^{2}} \\
\varepsilon ^{\left( 2\right) } &=&\varepsilon +\frac{1}{2}\Delta -\frac{1}{2%
}t+\frac{1}{2}\sqrt{\left( t+\Delta \right) ^{2}+8(t^{\prime })^{2}}\;\;\; 
\notag \\
\varepsilon ^{\left( 3\right) } &=&\varepsilon +t\;\;\;
\label{Isoscles1ElEnergies}
\end{eqnarray}%
$\allowbreak \allowbreak $In the equilateral limit we have $\Delta =0,$ $%
t^{\prime }=t$ and we recover $\varepsilon ^{\left( 1\right) }=\varepsilon
-2t$, \ $\varepsilon ^{\left( 2\right) }=\varepsilon +t,$ $\varepsilon
^{\left( 3\right) }=\varepsilon +t\ \ $ as found earlier for the equilateral
case. For an acute isosceles triangle (apical angle $\theta _{3}<90^{0}$, $%
\left| \vec{R}_{3}-\vec{R}_{1}\right| >$ $\left| \vec{R}_{2}-\vec{R}%
_{1}\right| $) state (1) is the one-electron ground orbital and $\varepsilon
^{\left( 3\right) }>\varepsilon ^{(2)}>\varepsilon ^{\left( 1\right) }$. If
the isoscles triangle is only slightly distorted from an equilateral
triangle then $\Delta $ and $\delta \equiv t^{\prime }-t$ are both small. In
this region we can Taylor-expand (\ref{Isoscles1ElEnergies}) giving an
energy gap $\allowbreak \allowbreak $

\begin{equation}
\varepsilon _{gap}\equiv \varepsilon ^{\left( 3\right) }-\varepsilon
^{\left( 2\right) }=-\frac{5}{6}\Delta -\frac{4}{3}\delta +O\left( \Delta
^{2},\delta ^{2},\Delta \delta \right) >0\;\;\;  \label{IsoscelesGap}
\end{equation}%
$\allowbreak $This will be the gap between the ground and first excited
orbital for independent electrons in an acute isosceles H$_{3}$, molecule.
The lowest three-electron states have the $\varepsilon ^{\left( 1\right) 
\text{ }}$orbital doubly ocupied, and for the acute triangle case the
groundstate has the $\varepsilon ^{\left( 2\right) }$ orbital singly
occupied. \ Equally, for an obtuse isosceles $H_{3}$ molecule the
groundstate involves $\varepsilon ^{(3)}$ rather than $\varepsilon ^{\left(
2\right) }$, and then (\ref{IsoscelesGap}) is negative, and the gap.is $%
\varepsilon _{g}=\varepsilon ^{\left( 2\right) }-\varepsilon ^{\left(
3\right) }\approx \frac{5}{6}\Delta +\frac{4}{3}\delta >0$.

The normalized one-electron eigenfunctions of (\ref{IsoscelesH1E1}) are,
with respect to s orbitals on the three protons as basis, 
\begin{eqnarray}
\left| v^{\left( 1\right) }\right\rangle  &=&\sqrt{N_{1}}\left( -\frac{1}{2}%
\frac{\varepsilon ^{\left( 2\right) }+\Delta -\varepsilon }{t^{\prime }},-%
\frac{1}{2}\frac{\varepsilon ^{\left( 2\right) }+\Delta -\varepsilon }{%
t^{\prime }},1\right)   \notag \\
\left| v^{\left( 2\right) }\right\rangle  &=&\sqrt{N_{2}}\left( -\frac{1}{2}%
\frac{\varepsilon ^{\left( 1\right) }+\Delta -\varepsilon }{t^{\prime }},-%
\frac{1}{2}\frac{\varepsilon ^{\left( 1\right) }+\Delta -\varepsilon }{%
t^{\prime }},1\right)   \notag \\
\left| v^{\left( 3\right) }\right\rangle  &=&\sqrt{N_{3}}(-1,1,0)
\label{IsoscelesEigenvectors}
\end{eqnarray}%
where the normalizing factors $N_{i}$ depend on the direct overlaps $\alpha
_{23}=\alpha _{13}$, $\alpha _{12}$. From (\ref{IsoscelesEigenvectors}) we
find that the one-electron states have large dipole moment vectors $\vec{d}$
for arbitrarily small isosceles distortions, even though the energy gap (\ref%
{IsoscelesGap}) is arbitrarily small. Specifically

\begin{eqnarray}
\vec{d}_{1} &\equiv &-\left| e\right| \left\langle v^{\left( 1\right)
}\right| \vec{r}\left| v^{\left( 1\right) }\right\rangle =\eta _{1}\hat{h}%
\;\;\;\;\;\vec{d}_{2}\equiv -\left| e\right| \left\langle v^{\left( 2\right)
}\right| \vec{r}\left| v^{\left( 2\right) }\right\rangle =\eta _{2}\hat{h},\;
\notag \\
\;\vec{d}_{3} &\equiv &-\left| e\right| \left\langle v^{\left( 3\right)
}\right| \vec{r}\left| v^{\left( 3\right) }\right\rangle =\eta _{3}\hat{h},\
\ \ \ \ \vec{d}_{23}\equiv -\left| e\right| \left\langle v^{\left( 2\right)
}\right| \vec{r}\left| v^{\left( 3\right) }\right\rangle =\eta _{23}\hat{b}
\label{Isosceles1BodyDipoleMatrix}
\end{eqnarray}%
where $\hat{b}$ is a unit vector pointing from proton 1 to proton 2 along
the base of the isosceles triangle, and $\hat{h}$ is a unit vector
perpendicular to $\hat{b}$ and directed towads the apex (proton 3): see Fig.
4 \ For independent electrons the coefficients take the nonzero values \ 
\begin{eqnarray*}
\eta _{1} &=&\frac{-\left| e\right| N_{1}}{3}\left( -\frac{1}{2}\left( \frac{%
\varepsilon ^{\left( 2\right) }+\Delta -\varepsilon }{t^{\prime }}\right)
^{2}\left( 1+\alpha \right) +2-\frac{1}{2}\frac{\varepsilon ^{\left(
2\right) }+\Delta -\varepsilon }{t^{\prime }}\alpha _{23}\right) h \\
\eta _{2} &=&\frac{-\left| e\right| N_{2}}{3}\left( -\frac{1}{2}\left( \frac{%
\varepsilon ^{\left( 1\right) }+\Delta -\varepsilon }{t^{\prime }}\right)
^{2}\left( 1+\alpha \right) +2-\frac{1}{2}\frac{\varepsilon ^{\left(
2\right) }+\Delta -\varepsilon }{t^{\prime }}\alpha _{23}\right) h \\
\eta _{3} &=&\frac{-\left| e\right| 2N_{3}\left( 1-\alpha \right) }{3}h \\
\eta _{23} &=&-\left| e\right| \sqrt{N_{2}N_{3}}\left( -\frac{1}{2}\frac{%
\varepsilon ^{\left( 1\right) }+\Delta -\varepsilon }{t^{\prime }}+\frac{1}{2%
}\alpha _{23\ \ \ \ }\right) b
\end{eqnarray*}%
where the $\left\{ \alpha _{ij}\right\} $ are overlap matrix elements
between neigboring atomic s functions, and the $\left\{ \sqrt{N_{i}}\right\} 
$ are normalizing factors for eigenfunctions $\left| \nu ^{\left( i\right)
}\right\rangle $.

From the one-body states $\left| \nu ^{\left( i\right) }\right\rangle $ we
construct two low-lying determinantal states $\left| B\right\rangle ,\left|
S\right\rangle $ of three independent electrons in the isoscleles $H_{3}$
triangle, each with two electron spins up ($\uparrow $) and one down ($%
\downarrow $) giving total spin angular momentum +$\hbar /2$:%
\begin{equation*}
\left| B\right\rangle =c_{1\uparrow }^{\dagger }c_{1_{\downarrow }}^{\dagger
}c_{2\uparrow }^{\dagger }\left| 0\right\rangle ,\ \ \left| S\right\rangle
=c_{1\uparrow }^{\dagger }c_{1_{\downarrow }}^{\dagger }c_{3\uparrow
}^{\dagger }\left| 0\right\rangle
\end{equation*}%
These states are separated by a small gap $\varepsilon _{gap}$ given by Eq (%
\ref{IsoscelesGap}).\ For the acute isosceles triangle, the state $\left|
B\right\rangle $ with an electric dipole moment pointing towards the base
(''B'') of the triangle is the groundstate, while the state $\left|
S\right\rangle $ with a dipole pointing to the apex or summit (''S'') is the
groundstate for the obtuse case. The dipole matrix elements for a single
molecule are 
\begin{eqnarray}
\vec{d}_{B} &\equiv &-\left| e\right| \left\langle B\right| \vec{r}_{1}+\vec{%
r}_{2}+\vec{r}_{3}\left| B\right\rangle =2\vec{d}_{1}+\vec{d}_{2}=-\left| 
\vec{d}_{B}\right| \hat{h},\   \notag \\
\ \vec{d}_{S} &=&-\left| e\right| \left\langle U\right| \vec{r}_{1}+\vec{r}%
_{2}+\vec{r}_{3}\left| U\right\rangle =2\vec{d}_{1}+\vec{d}_{3}=\left| \vec{d%
}_{S}\right| \hat{h}  \notag \\
\vec{d}_{BS} &\equiv &-\left| e\right| \left\langle B\right| \vec{r}_{1}+%
\vec{r}_{2}+\vec{r}_{3}\left| S\right\rangle =\vec{d}_{SB}=\left| \vec{d}%
_{UD}\right| \hat{b}  \label{IsoscelesH3DipoleElements}
\end{eqnarray}%
The vector directions (parallel to $\hat{h}$ or $\hat{b}$) of these matrix
elements stem from the mirror symmetry of the isosceles triangle, and remain
valid when we introduce the electron-electron interaction, thereby going
beyond our initial neglect of electron-electron interactions inside the H$%
_{3}$ triangle.

\subsection{Two parallel facing isosceles H$_{3}$ units, H$_{3}a$ and H$%
_{3}b $\ }

For small enough gap $\varepsilon _{g}$ we need only keep, as our
two-molecule product basis, the groundstate and lowest 3 excited
noninteracting states of the H$_{3}-$H$_{3}$ complex, namely $\left|
S\right\rangle _{a}\left| S\right\rangle _{b},\ \left| S\right\rangle
_{a}\left| B\right\rangle _{b},\ \ \left| B\right\rangle _{a}\left|
S\right\rangle _{b},\ \ \left| B\right\rangle _{a}\left| B\right\rangle _{b}$
in that order$.\ \ $In this basis the bi-molecular hamiltonian matrix
(relative to two isolated groundstate molecules), including the dipole
interaction $\delta H=e^{2}X_{a}.X_{b}/D^{3}$ between the molecules, is 
\begin{equation}
H_{ab}=\left( 
\begin{array}{cccc}
2\varepsilon _{g}+cD^{-3} & 0 & 0 & gD^{-3} \\ 
0 & \varepsilon _{g}+\beta D^{-3} & fD^{-3} & 0 \\ 
0 & fD^{-3} & \varepsilon _{g}+\beta D^{-3} & 0 \\ 
gD^{-3} & 0 & 0 & aD^{-3}%
\end{array}%
\right)  \label{HabIsosceles}
\end{equation}%
where 
\begin{equation}
a=\left| \vec{d}_{B}\right| ^{2}>0,\ \ \beta =\vec{d}_{B}.\vec{d}_{S}<0,\ \
c=\left| \vec{d}_{S}\right| ^{2}>0,\ \ f=g=\left| \vec{d}_{SB}\right| ^{2}>0
\label{CoeffsInHabIsosceles}
\end{equation}%
The hamiltonian (\ref{HabIsosceles}) is very similar to that for the
equilateral case in the broken-symmetry ''up'' ''down'' 1-electron basis:
see Eq (\ref{HPrimeab}). The zeros in the matrix (\ref{HabIsosceles}) come
from matrix elements such as%
\begin{equation*}
\left\langle B_{a}B_{a}\right| \vec{R}_{a}.\vec{R}_{b}\left|
B_{a}S_{b}\right\rangle =\vec{d}_{B}.\vec{d}_{BS}=\left| \vec{d}_{B}\right|
\ \left| \vec{d}_{BS}\right| \hat{h}.\hat{b}=0
\end{equation*}%
This zero arises mathematically from the orthogonality of the unit vectors $%
\hat{b}$ and $\hat{h}$ and is mandated physically by the mirror symmetry of
the isosceles molecule. This symmetry survives when our original neglect of
the intramolecular electron-electron interaction is relaxed. The form of Eq (%
\ref{HabIsosceles}) is therefore valid even with\ inclusion of
electron-electron interactions within each H$_{3}$ molecule, though the
coefficients $a,\beta ,c,...,g$ will be determined by the dipolar matrix
elements $\vec{d}$ with e-e interactions included. \ As for the equilateral
case, however, we expect that if the on-site e-e repulsion $U$ is too strong
compared with the hopping amplitude $t$, the states $\left| B\right\rangle $
and $\left| S\right\rangle $ may no longer be well-separated energetically
from the next excited state of the H$_{3}$.molecule, invalidating the
present analysis.

It is easily shown that in the limits $\Delta \rightarrow 0$, $t^{\prime
}\rightarrow t$, corresponding to the equilateral limit of isosceles
triangles, (\ref{HabIsosceles}) reduces to (\ref{HPrimeab}).

The eigen-energies of the interacting isosceles H$_{3}$-H$_{3}$ system
(measured from the groundstate of two isolated H$_{3}$ units), from
diagonalization of (\ref{HabIsosceles}), are as follows: $\ \ \ $%
\begin{equation}
\ \ E_{IV}=\frac{1}{2D^{3}}\left( c+a+2\varepsilon _{g}D^{3}+\sqrt{\left(
c-a+2\varepsilon _{g}D^{3}\right) ^{2}+4g^{2}}\right)  \label{E_IV_Isosc}
\end{equation}

\begin{equation}
E_{III}=\frac{1}{2D^{3}}\left( c+a+2\varepsilon _{g}D^{3}-\sqrt{\left(
c-a+2\varepsilon _{g}D^{3}\right) ^{2}+4g^{2}}\right)  \label{E_III_Isosc}
\end{equation}

\begin{equation}
\ E_{II}=\varepsilon _{g}+\frac{\beta +f}{D^{3}}  \label{E_II_Isosceles}
\end{equation}%
\begin{equation}
E_{I}\ =\varepsilon _{g}+\frac{\beta -f}{D^{3}}  \label{E_I_Isosceles}
\end{equation}

\subsection{Asymptotic regime of isosceles H$_{3}$- H$_{3}$ interaction:
conventional fixed dipolar, vdW and (excited) resonant interactions}

Since the (frozen) isosceles distortion away from an equilateral
configuration has introduced an energy gap $\varepsilon _{g}$ and a large
permanent electric dipole $\vec{d}_{B}$ in the H$_{3}$ groundstate, one
might initially expect that the groundstate H$_{3}-$H$_{3}$ interaction
would be a sum of a conventional attractive vdW interaction varying as $%
D^{-6}$, plus a fixed-dipolar interaction that varies as $D^{-3}$and that
can be attractive or repulsive. This is indeed the case in the truly
asymptotic regime defined by 
\begin{equation*}
D>D_{\max }\equiv \left( \left| \vec{d}_{\max }\right| ^{2}/\varepsilon
_{g}\right) ^{1/3}
\end{equation*}

where $\left| \vec{d}_{\max }\right| $ is the greatest of the dipole matrix
magnitudes from (\ref{IsoscelesH3DipoleElements}). In this regime we can
Taylor-expand the eigenenergies (\ref{E_IV_Isosc}), (\ref{E_III_Isosc})
giving the following energies, $E_{III}$, $E_{I},E_{II}$, $E_{IV}$ ordered
from lowest to highest:

\begin{equation}
E_{III}=aD^{-3}-\frac{g^{2}}{2\varepsilon _{g}}D^{-6}+\allowbreak O\left(
D^{-8}\right)  \label{E_III_isosceles_asy}
\end{equation}%
This state III is the groundstate of the H$_{3}$-H$_{3}$ dimer and exhibits
a conventional attractive $D^{-6}$ vdW interaction plus a repulsive $D^{-3}$
interaction between the fixed molecular dipoles (repulsive because we have
assumed that the two parallel-facing H$_{3}$ molecules have the same
alignment so that $a>0)$. The next-lowest energies are

\begin{equation}
\ E_{I}=\varepsilon _{g}+\frac{b-f}{D^{3}},\;\;\;E_{II}=\varepsilon _{g}+%
\frac{b+f}{D^{3}}.\   \label{E_I+2_isosceles_asy}
\end{equation}%
Since $\beta <0$ and $f>0$ (see (\ref{CoeffsInHabIsosceles})), state I with
energy $E_{I}$ is the first excited state of the H$_{3}$ dimer. Its energy
is just below $\varepsilon _{g},$ corresponding to its origin as a
superposition of two product states in each of which just one of the
molecules is excited. This is the attractive ''resonant interaction''
introduced already by Eisenschitz and London in 1930 \cite{Eisenschitz}.
Such resonant excited states depend on the two gaps having the same nonzero
value $\varepsilon _{g}$, and the coupled state is related to the concept of
excitons in condensed matter systems. Similar physics is important for
understanding intramolecular transport of photon energy in chromophores\cite%
{FRET2010}. \ State II has a similar physical origin but has a higher energy
than $E_{I}$ and can give repulsion rather than attraction.

The highest-lying moleular dimer state from (\ref{HabIsosceles}), labelled $%
IV$, has an energy close to twice the gap energy,

\begin{equation}
E_{IV}=2\varepsilon _{g}+O\left( D^{-3}\right)  \label{E_IV_isosceles_asy}
\end{equation}

\subsection{Subasymptotic regime of spooky $D^{-3}$ vdW interaction between
isosceles H$_{3}$ molecules}

We will now show that if the gap $\varepsilon _{g}$ is sufficiently small,
there is a significant sub-asymptotic spatial regime where the interaction
is always attractive and varies as $D^{-3}$, just as for the equilateral
case treated above. The relevent regime is 
\begin{equation}
A<D<D_{1}\equiv \left( \left| d_{\min }\right| ^{2}/\varepsilon _{g}\right)
^{1/3}  \label{DefnSubasyRegime}
\end{equation}%
where $\left| d_{\min }\right| $ is the least magnitude of the dipolar
matrix elements (\ref{IsoscelesH3DipoleElements}) and $A$ is the spatial
size of each molecule. In this regime the energies (\ref{E_I_Isosceles}) - (%
\ref{E_IV_Isosc}) can be written, with the lowest listed first and the
highest listed last:%
\begin{equation}
E_{I}\approx \varepsilon _{g}+\frac{\beta -f}{D^{3}}\;\;bimolecular%
\;groundstate,\;spooky\;attractive\;vdW  \label{E_I_subasy}
\end{equation}%
\begin{equation}
E_{II}\approx \varepsilon _{g}+\frac{\beta +f}{D^{3}}\;\;excited%
\;bimolecular\;state  \label{E_II_subasy}
\end{equation}%
\begin{equation}
E_{III}\approx \frac{1}{2D^{3}}\left( a+c-\sqrt{\left( a-c\right) ^{2}+4g^{2}%
}\right) \;excited\;bimolecular\;state  \label{E_III_subasy}
\end{equation}%
\begin{equation}
E_{IV}\approx 2\varepsilon _{g}+\frac{1}{2D^{3}}\left( a+c+\sqrt{\left(
a-c\right) ^{2}+4g^{2}}\right) >0\;\;highest\;bimolecular\;state
\label{E_IV_subasy}
\end{equation}

In this regime the terms proportional to $\varepsilon _{g}$ are small
compared with \ other terms in (\ref{E_I_subasy}), (\ref{E_II_subasy}) and (%
\ref{E_IV_subasy}) . By comparing (\ref{E_I_subasy}) with (\ref%
{E_I+2_isosceles_asy}) one learns that the spooky correlated state arises as
the small-gap limit of the attractive resonant interaction, but it is now
the bimolecular groundstate, and not an excited state as in the ususal
resonant interaction.

This subasymptotic regime does not exist for most molecules because, for
example with a gap $\varepsilon _{g}\approx $1 eV and a dipole moment $%
\left( 1%
\text{\AA}%
\right) \left( \left| e\right| \right) $ the outer limit of the
subasymptotic spatial regime from (\ref{DefnSubasyRegime}) is at $%
D_{1}\approx 2.5\;\text{\AA }$ which lies in the overlap region of
electronic clouds where the present approach is not valid. \ However with a
gap of $\varepsilon _{g}=$0.01 eV and the same dipole moment we find $%
D_{1}\approx 12\;$\AA .which leaves a viable sub-asymptotic region. Larger
dipole moments and smaller gaps will extend $D_{1}$ to larger values.

\section{Other molecules likely to have electronic degeneracy similar to H$%
_{3}$}

In the working above, the important features of a molecule leading to an
anomalous $D^{-3}$ vdW interaction with another such molecule in the dimer
groundstate, were that an individual molecule has two degerate electronic
groundstates that are coupled by the dipole operator.\ \ \ \ To achieve this
in a similar manner to the equilateral \ H$_{3}$ molecule studied above, we
propose that one should consider molecules with the following features

(i) Discrete $N$-fold rotational symmetry (and an odd number of electrons)
leading to two degenerate Bloch-type many-electron states $\Psi _{+},\Psi
_{-}=\Psi _{+}^{\ast }$ for each spin configuration, states that are a
time-reversed pair (complex conjugates), each of which is an eigenfunction
of the rotation operator $\hat{R}_{360^{0}/N}$.

(ii) Sufficiently \ large ratio $t/U$ of the hopping amplitude $t$ to
on-site electronic repulsion $U$ to ensure that the two symmetry-mandated
degenerate states are well-separated from higher states, validating our
very-small-basis CI treatment.

(iii) An odd $N$ value, $N=2n+1,\,\ \;\;=1,2,3,...$This is needed because
for the even case, $N=2n$, one can show that the relevant matrix element of
the dipole operator is zero. To prove this, note that for the spooky effect
we need a degenerate time-reversed pair of many-electron states $\Psi
_{1}\left( \vec{\xi}\right) \equiv \left\langle \vec{\xi}\right| \left|
+\right\rangle $,$\;\;\Psi _{2}\left( \vec{\xi}\right) =\left\langle \vec{\xi%
}\right| \left| -\right\rangle =\Psi _{1}^{\ast }\left( \vec{\xi}\right) $
that are both eigenstates of the rotation operator $\mathcal{\hat{R}}%
_{360/\left( 2n\right) }=\mathcal{\hat{R}}_{180/n}$. Here for an M-electron
molecule we have denoted 
\begin{equation*}
\vec{\xi}=\left( \vec{r}_{1},\vec{r}_{2},...,\vec{r}_{M}\right)
=(x_{1},y_{1},z_{1}:x_{2},y_{2},z_{2}:....x_{M},y_{M},z_{M})
\end{equation*}%
Since for even-order rotational symmetry $\Psi _{1}$ and $\Psi _{2}$ are
eigenstates of $\mathcal{\hat{R}}_{180/n}$, they are also eigenstates of $%
\left( \mathcal{\hat{R}}_{180/n}\right) ^{n}=\mathcal{\hat{R}}_{180}$:%
\begin{equation}
\mathcal{\hat{R}}_{180}\Psi _{1}=\theta _{1}\Psi _{1},\;\;\;\;\;\mathcal{%
\hat{R}}_{180}\Psi _{2}=\theta _{1}^{\ast }\Psi _{2}
\label{EigfOf180Rotation}
\end{equation}%
Further since two 180$^{0}$ rotations produce no net effect we have $\left( 
\mathcal{\hat{R}}_{180}\right) ^{2}=\hat{I}$ and so in (\ref%
{EigfOf180Rotation}) we have $\theta _{1}^{2}=1\Longrightarrow \,$\ $\theta
_{1}=1$ or $-1$

However the $180^{0}$ rotation operator $\mathcal{\hat{R}}_{180}$ is in fact
the spatial coordinate inversion operator in the plane perpendicular to the
rotational symmetry axis:%
\begin{equation*}
\mathcal{\hat{R}}_{180}\Psi \left( \vec{\xi}\right) \equiv \Psi \left( \vec{%
\xi}\,^{\prime }\right)
\end{equation*}%
where for an M-electron molecule 
\begin{equation*}
\vec{\xi}\,^{\prime
}=(-x_{1},-y_{1},z_{1}:-x_{2},-y_{2},z_{2}:....-x_{M},-y_{M},z_{M})
\end{equation*}

Then the matrix element of the molecular dipole operator between the two
time-reversed states is$\,\;$ 
\begin{eqnarray*}
\left\langle -\right| \vec{d}\left| +\right\rangle &=&-\left| e\right| \int
\Psi _{2}^{\ast }\left( \vec{\xi}\right) \left( \vec{r}_{1}+\vec{r}_{2}+...+%
\vec{r}_{M}\right) \Psi _{1}\left( \vec{\xi}\right) d^{3M}\xi \\
&=&-\left| e\right| \int \left( \Psi _{1}\left( \vec{\xi}\right) \right)
^{2}\left( \vec{r}_{1}+\vec{r}_{2}+...+\vec{r}_{M}\right) d^{3M}\xi
\end{eqnarray*}%
To obtain our spooky inter-molecule correlation for parallel-facing dimer
geometry, we need each molecule to exhibit non-zero cartesian components of
the dipole matrix element $\left\langle -\right| \vec{d}\left|
+\right\rangle $ in the xy plane perpendicular to the rotational symmetry
(z)\ axis. \ These components form a perpendicular dipole matrix element $%
\left\langle -\right| \vec{d}_{\perp }\left| +\right\rangle $

\begin{equation}
\left\langle -\right| \vec{d}_{\perp }\left| +\right\rangle \equiv d_{x}\hat{%
x}+d_{y}\hat{y}=-\left| e\right| \int \left( \Psi _{1}\left( \vec{\xi}%
\right) \right) ^{2}\left( \vec{r}_{1\perp }+\vec{r}_{2\perp }+...+\vec{r}%
_{M\perp }\right) d^{3M}\xi  \label{IntegralFordDipoleEleent}
\end{equation}%
But $\left( \Psi _{1}\left( \vec{\xi}\right) \right) ^{2}=\left( \theta
_{1}^{-1}\Psi _{1}\left( \vec{\xi}\,^{\prime }\right) \right) ^{2}=\theta
_{1}^{-2}\left( \Psi _{1}\left( \vec{\xi}\,^{\prime }\right) \right)
^{2}=1\left( \Psi _{1}\left( \vec{\xi}\,^{\prime }\right) \right) ^{2}$.
Thus $\Psi _{1}^{2}$ is even under the inversion $\vec{\xi}\rightarrow \vec{%
\xi}\,^{\prime }$of $\vec{r}_{i\perp }$ $(x_{i}$ and $y_{i}$) coordinates,
so that the integrand of (\ref{IntegralFordDipoleEleent}) is odd and the
integral vanishes.

The conclusion is then that we will not obtain a spooky $-D^{-3}$ vdW
interaction for a system with even (2n-fold) rotation symmetry, because the
implied inversion symmetry makes the needed dipolar coupling vanish. Rather
we should look only for molecules with odd rotational symmetry (3-fold,
5-fold,... )

A similar argument based on rotational symmetry shows that (for even or odd $%
N)$ the dipole moment in either of the two degenerate states is also zero:\ $%
\left\langle +\right| \vec{d}_{\perp }\left| +\right\rangle =\left\langle
-\right| \vec{d}_{\perp }\left| -\right\rangle $.$=\vec{0}$. :\ this was the
other matrix element needed to ensure that the correlation problem for more
general molecules is isomorphic to the H$_{3}$ problem treated above.

Thus to obtain a $D^{-3}$ vdW interaction in a similar fashion to that
obtained for a pair of H$_{3}$ molecules above, we need to search for
molecules that have discrete odd rotational symmetry. Regular (2n+1)-gons
satisfy this when $n=1,2,3,.....$. So do a large number of cluster
structures: for two examples see Fig 5. \ Of course, this combination of
properties may not be the only way to achieve spooky vdW interactions, but
it does suggest one class of molecules to explore.

\begin{figure}
\includegraphics[width=1.0\linewidth]{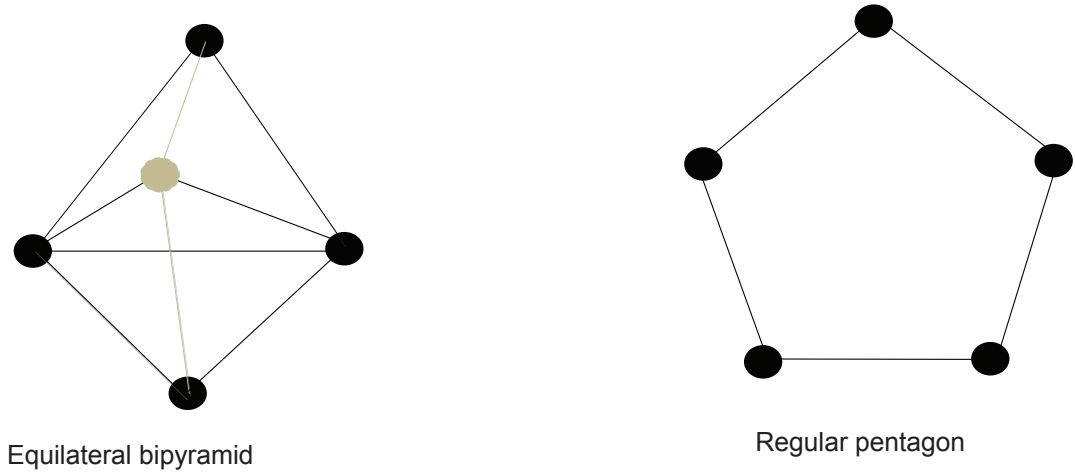}%
\caption{Two geometries conducive to spooky vdW forces}
\label{Fig5}
\end{figure}

Furthermore, our numerical studies of equilateral H$_{3}$ showed that the
hopping of electrons between sites needs to dominate over on-site coulomb
repulsion in order for the two-fold degenerate groundstates states to be
well separated from the first exicted state, as required for the validity of
our limited-CI analysis of the inter-molecular interaction. These
requirements suggest that clusters of metal atoms might be suitable
candidates.

As an example of the importance of breaking inversion symmetry via an odd
rotational symmetry, we also studied the square H$_{4}$ molecule, which has
discrete (90$^{0}$) rotational symmetry and therefore has inversion
symmetry, unlike equilateral H$_{3}$. We found no anomalous $D^{-3}$ vdW
interaction between two H$_{4}$ molecules, though an always-attractive $%
D^{-5}$ vdW interaction may be possible via spooky coupled quadrupolar
fluctuations. This work will be described elsewhere.

\section{ Effects of nuclear motion - metal atom clusters.}

The Jahn-Teller-distorted state of our target molecules should have a
permanent electric dipole, if the geometrical distortion is static, and
indeed many quantum chemical calculations have predicted large static dipole
moments for small metal-atom clusters (see \cite{NaClusterTh2011} for
example). Experimentally these large dipoles are not seen, even at cryogenic
temperatures where presumably thermal motions of the nuclei are less
relevant.\cite{AreNaClustersMet}. This puzzle has attracted considerable
recent attention and the current explanations involve nuclear motion
(pseudo-rotation) whereby a small change in the distortion can produce a
very different, sometimes opposite, dipole. Recent work \ within the
Born-Oppenheimer approximation has suggested that the shapes of these
clusters vary quite strongly over time \cite{MDalkaliMetalClustersM3} at
room temperature with both pseudorotation and shape-inversion present, while
at 20K only the pseudorotation is present. At low temperatures these nuclear
motions would be quantal, and when electronic degeneracies exist the motions
may be anharmonic because of the conical intersection physics. Indeed a
proper treatment will require a description of the coupled vibronic motions
of the electrons and nuclei, and in this regime one may speak of the dymamic
Jahn-teller effect (\cite{HamReDynJahnTeller}). We tentatively suggest that
the presence of a second such molecular cluster within the sub-asymptotic
regime (see (\ref{DefnSubasyRegime})) can significantly affect these
vibronic phenomena, leading to a coupling of the vibrational as well as the
electronic motions of both molecules. If, during the course of these coupled
motions at fixed intermolecular separation $D$, the instantaneous electronic
gap $\varepsilon _{g}$ satisfies the subasymptotic criterion (\ref%
{DefnSubasyRegime}) with significant probability, then one expects to see an
anomalous $D^{-3}$ vdW interaction similar to that discussed above for the
sub-asymptotic regime of isosceles H$_{3}$.

\section{Summary and Discussion}

We first studied an idealized system consisting of two interacting
equilateral $H_{3}$ molecules separated by distance $D$, each molecule
having frozen nuclear positions. We showed that this toy model exhibits
''spooky''corelations between the fluctuating molecular electric dipole
moments. These correlations do not decay with increasing intermolecular
separation $D$, leading to a van der Waals interaction energy falling off as 
$-D^{-3}$ $\ $(see Eq (\ref{Phie}). rather than the conventional $-D^{-6}$.
This interaction occurs in the groundstate of the molecular dimer, and can
be regarded as the zero-gap limit of the so-called ''resonant interaction'',
although the latter occurs only in an excited state when each molecule has a
finite gap. This physics is only possible where each molecule has degenerate
groundstates coupled by the electric dipole operator. We suggested that a
class of molecules worth exploring for the existence of such a spooky vdW
interaction are those with an odd number of electrons, with discrete ($2n+1)$%
-fold rotational symmetry about an axis in its ideal maximally symmetric
configuration, and that therefore break inversion symmetry in that
configuration. Equilateral H$_{3}$ satisfies these criteria.

We also considered a small static Jahn-Teller distortion, leading to an isosceles H$_3$ molecule.  
This system has a finite electronic gap and so two such molecules exhibit a conventional $-D^{-6}$ dispersion
interaction as $D\rightarrow \infty$.  Using  a limited CI model that can be solved analytically,  we also found however that there can be an intermediate range
 of separations $A < D < D_1$ (see Eqs \ref{DefnSubasyRegime} and \ref{E_I_subasy})  where the interaction energy is of form $-A_3 D^{-3.}$, similarly to the undistorted equilateral case. 

One of the most successful approaches for computation of dispersion interactions between "normal" 
gapped molecules is  Symmetry-Adapted Perturbation Theory (SAPT, \cite%
{PertThApprVdwComplexes}). It is clear that non-degenerate (single-determinant) perturbation theory, including SAPT,
 cannot work for the vdW interaction when there is
a strictly degenerate groundstate on each interacting molecule (as occurs
for the toy equilateral H$_{3}$ molecules studied in this paper)\ .  For
systems with fairly small but finite gaps such as cases with static Jahn-Teller distortion, 
there certainly exists work within
SAPT on the $C_{6}$ vdW coefficient of the $D^{-6}$ iteraction, applicable
at fully asymptotic separations,- e.g. \cite{Patkowski:07B}. The question
arises, though, whether SAPT can obtain the $D^{-3}$ interaction found in
the present work above for slightly distorted (isosceles) H$_{3}$ molecules
in the sub-asymptotic regime of intermediate separations $D$ (see Secs. 6A-C)
This may be possible because the
electronic groundstate of isosceles H$_{3}$ is no longer strictly
degenerate,\ but perhaps it requires a high order of pertubation theory in
SAPT. We hope to investigate this in a future publication. \ One needs also
to bear in mind that nuclear motions may make such statically distorted
calculations less relevant: see below.

H$_{3}$ is unfortunately unstable, but there exist odd-$N$ clusters of metal
atoms and other more rigid structures (e.g. \cite{BigMolecEquilatSymm}) with
equilateral (three-fold, 120$^{0}$) or five-fold, or seven-fold....
rotational symmetry etc., and broken inversion symmetry. For two examples of
such structures see Fig 5. These may be more stable against geometric
(Jahn-Teller) distortions than  H$_{3}$, and so may exhibit
dipole-allowed transistions between nearly- degenerate groundstates. Thus
they may be candidates for a $D^{-3}$ dispersion interaction, at least at
sub-asymptotic separations (see the criterion of Eq (\ref{DefnSubasyRegime})
for the sub-asymptotic regime). Furthernore, clusters of metal atoms are
promising because they tend to have large hopping amplitude 
$t$ between
neighboring atoms, as evidenced by their ability to form highly conductive
solids upon aggregation. These small cluster systems will tend to
Jahn-Teller distort to produce large fixed dipoles, but such dipoles are not
seen in experiments on metal atom clusters. The likely explanation lies in
small nuclear motions that induce large fluctuating dipoles
(''pdeudo-rotation''). \ We suggest that for dimers of these clusters, our
novel vdW attraction would be mediated by coupled vibronic motions of both
electrons and nuclei on both of the interacting molecules. \ If the
molecules pass near the high-symmetry, electronically degenerate states
sufficienly often during these combined motions, an appreciable weight of %
$D^{-3}$ vdW interaction should be observable. The analysis in \cite{Allen05} and in the recent work of Requist, Tandetzky and Gross \cite{Requist16} may be useful in analyzing this situation.

\section{Acknowledgments.}

We acknowledge conversations with Bogumil Jeziorski, Tim Gould, Georg Jansen and Ryan Requist.


%

\end{document}